\documentclass[final,5p,times,twocolumn]{elsarticle}
\usepackage[utf8]{inputenc}
\usepackage{amsmath}
\usepackage{amssymb}
\usepackage{graphicx}
\usepackage{multirow}
\usepackage{multicol}

\DeclareGraphicsExtensions{.png,.pdf,.jpg} 
\usepackage{textcomp} 

\usepackage[
pdfauthor={Benoit Clément},
pdftitle={Spatial resolution determination of a position sensitive ultra-cold neutron detector},
pdfkeywords={},
colorlinks={true},
linkcolor=red,
citecolor=green,
urlcolor=cyan,
pdfstartview={FitH}
]{hyperref}

\begin{document}
\begin{frontmatter}
 
\title{Spatial resolution determination of a position sensitive ultra-cold neutron detector}
\author[LPSC]{B.~Clément\corref{cor1}}\ead{bclement@lpsc.in2p3.fr}
\author[IPHC,GSI]{L.~Gesson}
\author[ILL]{T.~Jenke}
\author[ILL]{V.V.~Nesvizhevsky}
\author[LPSC]{G.~Pignol}
\author[ILL,LPSC]{S.~Roccia}
\author[LPSC]{J.-P.\,Scordillis}
\cortext[cor1]{Corresponding author}
\address[LPSC]{Univ. Grenoble Alpes, CNRS, Grenoble INP*, LPSC-IN2P3, 38000 Grenoble, France
* Institute of Engineering Univ. Grenoble Alpes}
\address[ILL]{ILL, Institut Laue Langevin, 71 avenue des Martyrs, CS 20156, 38042 Grenoble Cedex 9, France}
\address[IPHC]{IPHC, UMR 7178, Université de Strasbourg, CNRS, 67037 Strasbourg, France}
\address[GSI]{GSI Helmholtzzentrum f\"ur Schwerionenforschung GmbH, Planckstrasse 1, 64291 Darmstadt, Germany}
\begin{abstract}
The study of the properties of the quantum states of bouncing neutrons requires position sensitive detection with micro-metric spatial resolution. 
The UCNBoX detector relies on Charge Coupled Devices (CCD) coated with a thin boron-10 conversion layer to detect neutron hits.
In this paper, we present an original experimental method to determine the spatial resolution of this device using micrometric masks. The observed resolution is $2.0\pm0.3~\mu$m.
\end{abstract}

\end{frontmatter}

\section{Introduction}
When a neutron wavelength becomes larger than the inter-atomic spacing it can scatter coherently on atoms. This phenomenon can be described by an effective potential called Fermi potential which translates into a critical velocity of the order of a few m.s$^{-1}$ for most materials. Neutrons with velocities smaller than the critical velocity are reflected at any angle of incidence. Such neutrons are called \emph{ultra-cold} neutrons (UCNs).

Ultra-cold neutrons are used in several particle physics experiments probing fundamental interaction~\cite{Dubbers2011}. In particular, UCNs are one of the very few probes to study gravity in a quantum context. The vertical motion of a bouncing neutron on the surface of a flat, horizontal mirror, exhibits quantum behavior~\cite{Nesvizhevsky2002,Nesvizhevsky2010,Cronenberg2018,Pignol2015}.
This system has stationary quantum states with discrete energy and wave functions whose spatial extension is governed by the parameter $z_0 = (\hbar^2/2m_n^2g)^{1/3} \approx 5.8\, \mu$m. 
Resolving the spatial structure of the wave functions requires a sensitivity to the neutron vertical position at the micron scale.

In section 2 of this article we discuss quickly the design of a silicon based pixelated neutron detectors. The third section focuses on the identification of the neutron hits and section 4 discusses the estimation of the resolution of the sensor using thin wire masks. 

\section{UCN pixelated silicon detectors}
\subsection{UCN detection}
To be sensitive to neutrons, solid state detectors need a conversion layer absorbing the neutrons to create energetic charged particles. 
First generation of gravitational neutron quantum states experiments employed plastic (CR-39) nuclear-track detectors to detect charged particles~\cite{Nesvizhevsky2000}. Such detectors lack real-time readout capability as they need a chemical etching process to reveal the impact of a charge particle and then read out. Therefore the possibility of using active sensors is being investigated, in particular pixelated silicon sensors such as CCD (charge Coupled Device) and CMOS (Complementary Metal-Oxide-Semiconductor) sensors. Sub-pixel resolution is achieved with the barycenter method of adjacent pixels. 

Only four stable nuclei can convert slow neutrons to charged particles: $^3$He, $^6$Li, $^{10}$B and $^{235}$U. For a solid detector to be sensitive to UCN, the conversion layer must fulfill several constraints: to reflect UCNs as little as possible (Fermi potential close to 0), to have a large absorption efficiency (large capture cross section).
The most practical candidate is $^{10}$B which represents 20\% of natural boron. Neutrons are converted by the reaction:
$\textrm{n} + {}^{10}\textrm{B}\rightarrow {}^7\textrm{Li} +\alpha    $
with a 1.5~MeV $\alpha$ particle and a 0.8\,MeV lithium ion in 94\% of the cases. Either of these particles can be detected. Natural boron has a critical velocity of $5.9$\,m.s$^{-1}$ whereas it is close to $0$\,m.s$^{-1}$ for enriched boron with more than 97\% $^{10}$B. Enriched boron must therefore be used to produce the conversion layer to maximize the conversion efficiency.

To determine the resolution the sensor is exposed to a neutron flux partly hidden by a mask. We want the edge of the shadow to correspond to the resolution, so the further the mask from the sensor, the less divergent the neutron beam must be. 

Using a collimated cold neutrons beam allows to place the mask a few hundred microns away from the sensor surface. This is the case in~\cite{Kawasaki2010} where a gadolinium mask casting a shadow on a commercial back-illuminated CCD sensor (with a pitch of $24 \, \mu$m) coated with a$^{10}$B led to a $2.9\pm0.1\,\mu$m resolution. A similar technique was used in~\cite{Kamiya2020} resulting into a $4.1\pm0.2\,\mu$m resolution for a $17\,\mu$m pitch boron coated CMOS sensor.

As UCNs will mostly interact in the front of the conversion layer, where cold neutrons will have a uniform interaction along the full layer. As a consequence the charged particles from UCNs conversion are expected to have a longer path through the conversion layer, thus degrading the resolution. It is therefore desirable to determine resolution with UCNs. Since UCN "beams" usually have a large angular distribution, the mask must be in contact of the sensor surface. 
In \cite{Jakubek2009,Jakubek2009b} a CCD device coated with $^6$LiF was exposed to ultra-cold neutrons. A spatial resolution of $2.3 \, \mu$m for a pixel pitch $55 \, \mu$m was claimed using the edge of a nickel foil on the sensor surface as a mask. With another CMOS sensor~\cite{Kuk2019}, kapton masks were used to achieve the same goal, albeit with poorer results. Both used large surface sensors and the technique might be unsuitable for smaller sensors such as used in the UCNBoX detector. We propose a solution to estimate the resolution using UCNs and thin wire masks as a relatively simple experimental setup compensated with more sophisticated analysis techniques.

\subsection{The UCNBoX detector}

The UCNBoX was designed to measure the wave functions of the bouncing neutrons in the GRANIT instrument~\cite{Roulier2015}. 
The spatial extension of the searched signal is less than a hundred microns so we chose a CCD sensor with a rectangular (rather than squared) active area. The CCD used is a Hamamatsu S11071-1106N sensor with $2048\times64$ pixels with a 14\,$\mu$m pitch. The sensor has no entrance window but the active area recedes 300$~\mu$m below the edge of the silicon bulk. This combined with the small height of the sensor makes it impossible to use a macroscopic mask as in previous experiments.

\begin{figure}[hh]
\begin{center}
\includegraphics[width=0.45\textwidth]{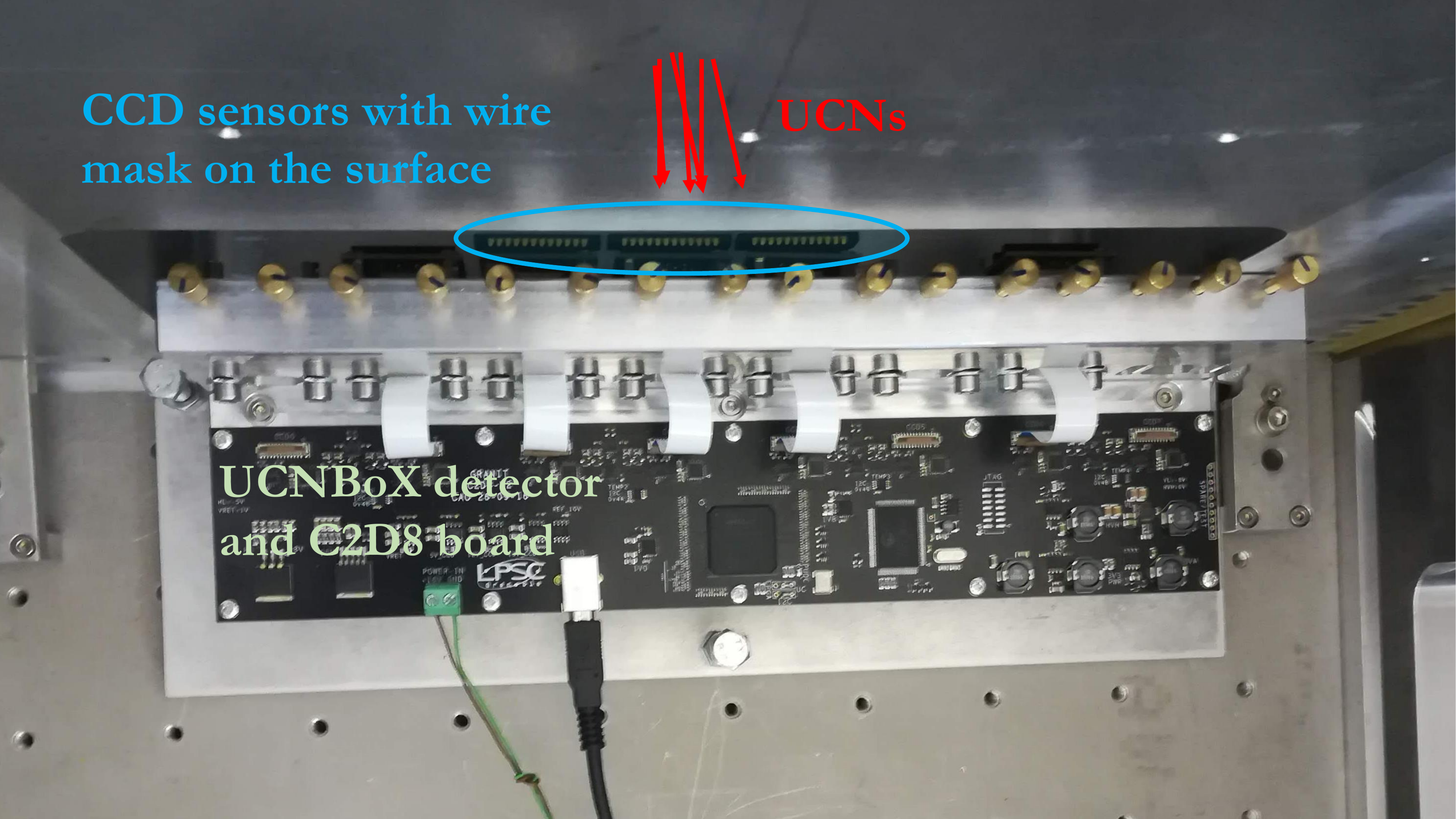}
\caption[]{The experimental setup with the UCNBoX detector loaded with 3 coated sensors facing the PF2 Test UCN beam.}
\label{fig-box}
\end{center}
\end{figure}

The UCNBoX uses up to eight sensors to cover the 30\,cm width of the experiment. Each sensor can be mechanically aligned with respect to its neighbors. A dedicated electronics~\cite{Bourrion2018} allow to read all eight sensors sequentially and transmit the data to a computer through a USB connection. A picture of the detector is given in figure~\ref{fig-box}.
The conversion layer~\cite{Clement19} consists on a thin 200\,nm layer of enriched boron (96\% of $^{10}$B) in between two 20\,nm layers of titanium and nickel. The external titanium layer protects boron against oxidization and limits neutron reflection. The second layer aims at improving the adhesion of the boron on the silicon substrate. This structure has been optimized to offer a good and uniform efficiency over a large velocity spectrum. 

\section{Neutron hit reconstruction}
\label{hitreco}
\begin{figure}[h]
\begin{center}
\includegraphics[width=0.45\textwidth]{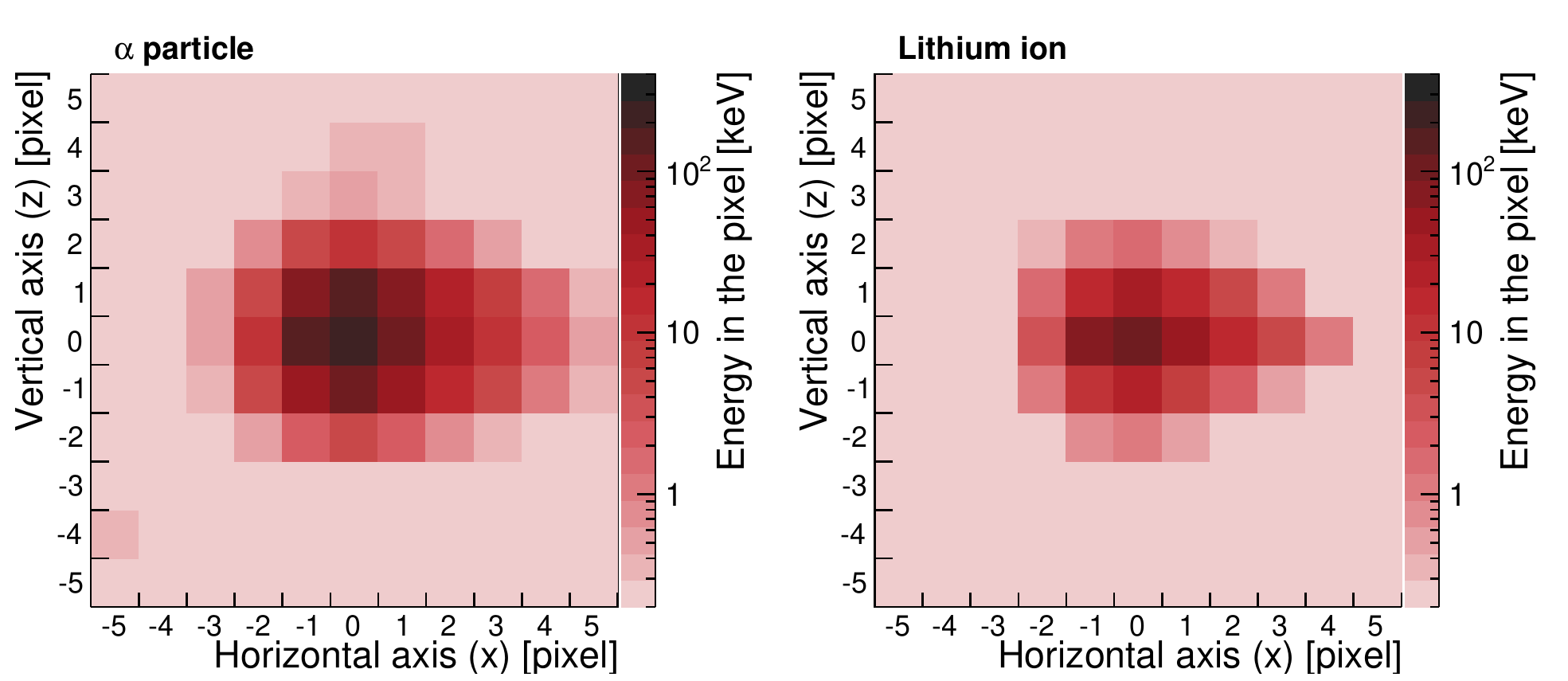}
\caption[]{Average energy deposition in each pixel for an $\alpha$ particle (left) and a lithium ion (right) from neutron capture on boron.}
\label{fig-cluster}
\end{center}
\end{figure}

The range in silicon of the ions produced in the neutron capture on boron does not exceed a few microns with an even smaller lateral straggling. The produced charge will drift within the bulk silicon and several pixels will be hit as shown in figure~\ref{fig-cluster}. The position could be reconstructed using a weighted average over a cluster of pixels. The fraction of the charge measured in square rings around the center is given in table~\ref{tab-adcfrac}. 

\begin{table}[!bth]
\label{tab-adcfrac}
\begin{center}
\vspace{-1cm}
\begin{small}
\caption[]{ Charge fraction in various size of square clusters around the seed pixel for 1.5 MeV $\alpha$ particle and 0.8\,MeV lithium ion from neutron capture on boron.}
\begin{tabular}{|c|cc||c|cc|} \hline
Size & $\alpha$ & $^7$Li & Size & $\alpha$ & $^7$Li \\
\hline
$1\times 1$       & 18.0\%  & 29.5\% & $3\times 3$       & 84.6\%  & 84.7\%\\
$5\times 5$       & 95.8\%  & 94.3\% & $7\times 7$       & 98.3\%  & 96.9\%\\
$9\times 9$       & 99.2\%  & 98.4\% & $11\times 11$     & 99.9\%  & 99.2\%\\
\hline
\end{tabular}
\end{small}
\end{center}
\end{table}

From this a dedicated clustering algorithm follows. For each image, a $3\times3$ sliding window is applied to determine seed clusters whose total ADC count exceed a given threshold. After this first clustering, all the pixels not falling into an $11\times11$ pixels cluster around the seed are used to estimate noise. The charge average of unclustered pixels is computed for each column and subtracted to the charge in each pixel of that column. 
The clustering procedure is then run again on the background subtracted image.
Each cluster is afterward characterized by three variables: the reconstructed horizontal $x_r$ and vertical $z_r$ positions and the sum of ADC counts in the cluster. 

\section{Determination of the spatial resolution}
\subsection{Experimental setup}
A measurement was done using UCNs from the PF2 Test beam at ILL over 10 days. A few  pieces  of 10\,$\mu$m diameter nickel wires about 1\,mm long were placed on the surface of three coated CCD sensors. Each wire acts as a mask for neutrons. We chose nickel for its high critical velocity of 6.9\,m.s$^{-1}$. Nevertheless as the PF2 beam velocity spectrum is centered at 7\,m.s$^{-1}$, we expect that 20 to 40\% of the UCN flux can go through the mask.
The detector UCNBoX with the three sensors installed is placed in a vacuum chamber, facing the UCN beam as shown in figure~\ref{fig-box}.

A similar experiment was done using $6\,\mu$m tungsten wire and $\alpha$ particles of energy $E\approx 1.4$\,MeV (obtained using 5.48\,MeV $\alpha$ from ${}^{241}$Am decay going through 12\,$\mu$m of aluminum and 19\,mm of air). These data allow to estimate the maximal resolution without the effect of the conversion layer.

We identify the regions of the sensor containing pieces of wire and cut them into small datasets of neutron position $(x_r^{i},z_r^{i})$ with a few thousand points each, corresponding to length of wire of 150 to 250\,$\mu$m. Twelve such dataset are analyzed independently and combined,  corresponding to a total length of mask of $2.5$\,mm surrounded by 0.5 to 0.7 neutron per $\mu$m$^2$.

\subsection{Reconstruction of the mask}
As the position of each mask is not precisely known, we will use the data to identify each mask and reconstruct its center. As each piece of wire might not be exactly straight, we used a third order polynomial for the mask shape. 

The reconstruction of the mask position from the data needs to build an estimator of vacuity (or inverse density). Such an estimator associate to each neutron hit a value that will quantify the scarcity of neutrons around it. One will expect this estimator to be large and uniform under the mask is and smaller and uniform outside with a transition between the two regions.
Such an estimator can be obtained through the Voronoi tessellation~\cite{voronoi08a,voronoi08b}. Starting with a set of 2D points, one first compute the optimal triangular mesh using Delaunay triangulation~\cite{delaunay34}. The the dual graph of the Delaunay triangulation if the Voronoi tessellation. Each point of the original set $(x_r^{i},z_r^{i})$ is contained in exactly one cell which can be viewed as the region of influence of the point. The cell area $a(x_r^{i},z_r^{i})$ is an estimator of the vacuity surrounding it, the cells being larger in less populated regions. Delaunay triangulation and Voronoi tessellation are standard tools for meshing in finite elements numerical methods and therefore free implementation of the algorithms are easily available such as  Triangle~\cite{shewchuk96b} which was used in this study. In addition specific code was developed to identify the Voronoi cells and compute their area.
We build an ad-hoc model of density:
\begin{equation}
A(x,y)=l_0+(l_1-l_0)\exp\left[-\left(\frac {d(x_r,z_r,u,v,w,c)}{s}\right)^4\right]
\end{equation}
where $l_1$ and $l_0$ describe the average cell area in the regions with and without mask, $d(x_r,z_r,u,v,w,c)$ is the distance of the point $(x_r,z_r)$ to the polynomial curve $z=ux^3+vx^2+wx+c$ which will define the mask center, and $s$ will quantify the width of the mask. The power $4$ in the exponential is purely ad-hoc and allows for a flatter top and a better data-to-model agreement than a Gaussian profile. Nevertheless the Gaussian profile (power 2) has been tried and leads to similar results. The difference in the final resolution is included in the systematic errors.

\begin{figure}[hh]
 \begin{center}
 \vspace{-0.5cm} 
 \includegraphics[width=0.5\textwidth]{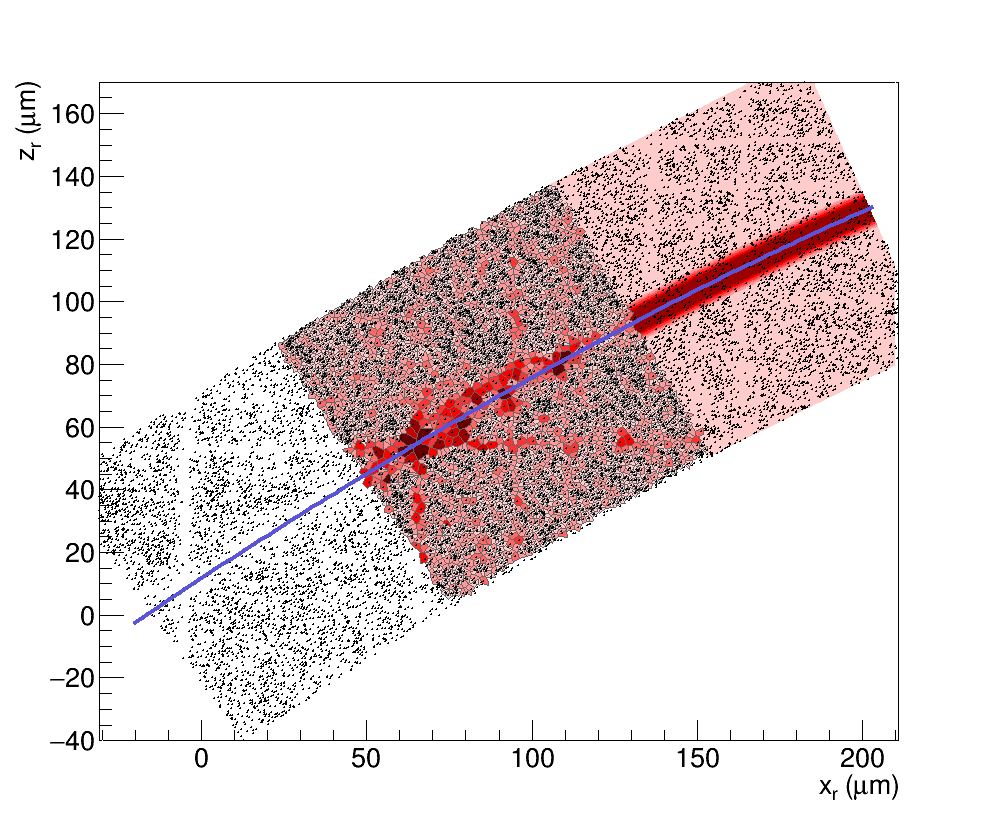}
 \vspace{-0.5cm} \caption[]{The steps of the analysis - for the same piece of nickel wire, each dot indicates a reconstructed UCN position : the raw data (left part), the Voronoi tessellation (center part), the 2D fit to the cell surface (right part) and finally the wire direction from the previous fit (blue line).}
 \label{fig-reco}
 \end{center}
 \end{figure}

We perform a $\chi^2$ fit of the 2D graph $a(x_r^{i},z_r^{i})$ to the model to extract the wire center position. The main steps of the analysis chain on various parts of a single mask are illustrated on figure~\ref{fig-reco}, from left to right: the neutron hits, the Voronoi tessellation, the 2D fit of the vacuity model and as a blue curve the final fitted wire center position.  

\subsection{Estimation of the resolution}
\begin{figure}[hh]
 \begin{center}
 \includegraphics[width=0.4\textwidth]{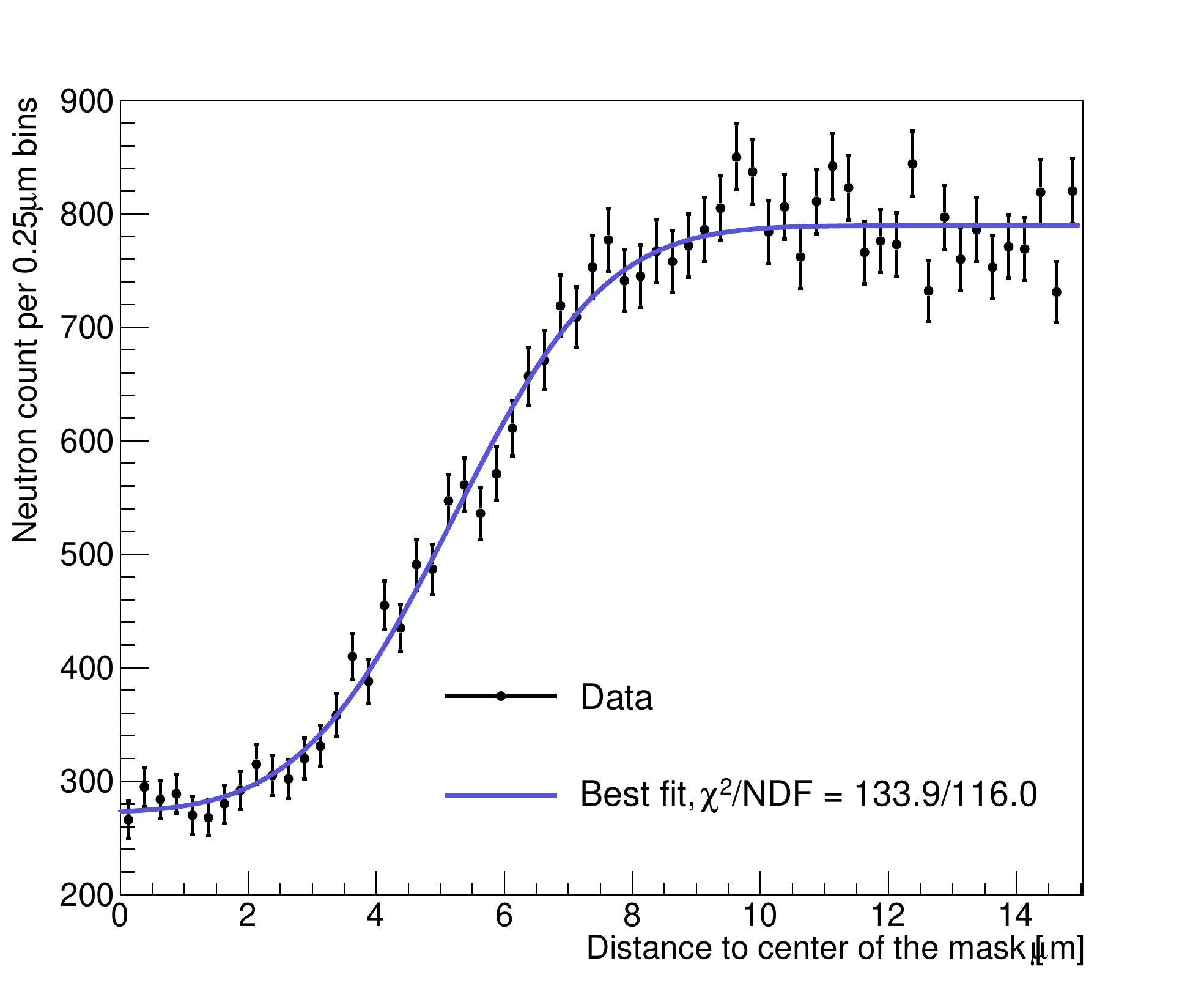}
 \vspace{-0.5cm} \caption[]{Distance of the neutron impacts to the wires, combining all available data. The fit gives access to the wire diameter and the raw resolution.}
 \label{fit-final}
 \end{center}
 \end{figure}

Once the central position of the mask has been found, we compute the distance of each reconstructed neutron to the mask center and build an histogram combining all datasets. Finally, this histogram is fitted with a error function:
  \begin{equation}
 f(d) = \frac{A}{2}\left(1+\textrm{erf}\left(\frac{d-\frac{w}{2}}{\sigma}\right)\right) +\frac{B}{2}\left(1-\textrm{erf}\left(\frac{d-\frac{w}{2}}{\sigma}\right)\right)
 \end{equation}
 where $w$ is the wire diameter, $\sigma$ the standard deviation of the line spread function, $A$ and $B$ the amplitudes outside and under the mask as shown on figure~\ref{fit-final}.
 The fitted wire diameter is $w=10.4\pm0.2\,\mu$m, in agreement with the expected value. The standard deviation gives the raw resolution $\sigma = 2.6\pm 0.1\,\mu$m where the error accounts only for the statistical effects. 
 
 The raw resolution $\sigma$ is but an upper bound on the true resolution $\sigma_{UCN}$. Because of the circular shape of the mask and the angular dispersion of the incoming neutrons, one does not expect a sharp edge even for a perfect resolution. We have to rely on Monte-Carlo simulation to take these effects into account. The simulation includes the generation of UCNs according to PF2 velocity spectrum and the interaction of UCNs with the mask, including higher velocity UCNs going through the mask. Finally the hits are shifted according to a Gaussian resolution. The simulated masks shape and length and the off-mask neutron density are those observed in the data. The simulation produces 2D neutron hits that are then forwarded through the same analysis chain as real data. This allows to map the raw resolution to the one injected into the simulation and estimate the true resolution of the sensor. The simulation is also used to estimate the systematic error induced by the analysis procedure itself comparing the reconstructed wire direction to the simulated ones. The total systematic error is 0.3$\,\mu$m, which dominates the total uncertainty.We conclude that the measured resolution of our sensors is 
 $\sigma_{UCN} = 2.0\pm 0.3\,\mu\textrm{m}$. Similar measurements with $\alpha$ particles leads to a resolution $\sigma_\alpha = 1.0\pm 0.2\,\mu \textrm{m}$ which is probably the lowest resolution achievable with this device. 

Unfortunately, horizontal and vertical empty regions appear on the data as seen on the left part of figure~\ref{fig-reco}). These lines are in fact several overlapping squares of $11\times11$ pixels corresponding to the clustering size used in the reconstruction. These structures were not observed in earlier measurement with the same sensors two years before. After analyzing the raw data before cluster reconstruction, it was possible to attribute this effect to noisy pixels that tend to attract the reconstructed clusters. Coming from noise fluctuations it is a stochastic effect that cannot be corrected. Yet we were able to be simulate by Monte-Carlo the effect of such noisy pixels and reproduce the observed pattern. The same simulation shows that this effect degrades the resolution which should be closer to 1.5\,$\mu$m. As it was not seen before using the same sensors, we concluded that it came from aging and/or bad storage of the sensors (the boron layers were more than four years old).

\section{Conclusion}
\label{conclusion}
We proposed a original method to determine the spatial resolution of position sensitive ultra-cold neutron detectors. It consists of using micrometric masks composed of small pieces ($\sim 1$\,mm long) of thin diameter ($10\,\mu$m or less) nickel wires on the surface of the sensor. The position of each mask is identified using an adjustment to the area of the Voronoi tessellation cell associated to each neutron hit. Finally the resolution is derived from the distribution of the distance of neutron hits to the masks.
A Monte-Carlo simulation is used to correct the measured raw resolution for the shape of the wire and the angular distribution of UCNs. In addition such simulation allows to check the systematic uncertainty of the reconstruction method.
Applying this technique to the CCD sensors of the UCNBoX detector, we demonstrate a resolution of $2.0\pm 0.3\,\mu$m.
This result is obtained on a old, noisy sensor and we estimate that a fresh sensor would exhibit a better resolution of the order of $1.5\,\mu$m.
\section*{Acknowledgments}
The authors would like to thank A.~Lacoste, Y.~Xi and A.~Bes at LPSC for the production on the conversion layer as well as T. Brenner for his tremendous support on the experiment at Institut Laue-Langevin.
\bibliographystyle{ieeetr}
\bibliography{boronreso} 
\end{document}